\documentclass[aps,prl,twocolumn,showpacs]{revtex4-1}
\usepackage{amsmath}
\usepackage{amsfonts}
\usepackage{amssymb}
\usepackage{graphicx}
\usepackage{graphicx}
\usepackage{epstopdf}
\usepackage[caption=false]{subfig}
\usepackage{subfig}
\usepackage{ulem}
\usepackage{mathtools}
\usepackage[autostyle]{csquotes}
\usepackage{color}
\usepackage{csquotes}
\usepackage[utf8]{inputenc}
\usepackage{fontenc}
\usepackage{lipsum}
\usepackage[english]{babel}
\usepackage{geometry}
\usepackage{marginnote}
  \makeatletter
    \renewcommand\@make@capt@title[2]{%
     \@ifx@empty\float@link{\@firstofone}{\expandafter\href\expandafter{\float@link}}%
      {\textbf{#1}}\@caption@fignum@sep#2\quad}%
    \makeatother
   
\makeatletter
\renewcommand{\fnum@figure}{\textbf{Figure~\thefigure}}
\makeatother  

\begin{document}

\title{Laminar-Turbulent Transition in Raman Fiber Lasers: A First Passage Statistics Based Analysis}

\author{Amit K Chattopadhyay}
\email{Corresponding author: a.k.chattopadhyay@aston.ac.uk}
\affiliation{                    
Aston University, Mathematics, Birmingham B4 7ET, UK}

\author{Diar Nasiev}
\affiliation{                    
Aston University, Mathematics, Birmingham B4 7ET, UK}

\author{Srikanth Sugavanam}
\affiliation{Aston Institute of Photonics Technologies, Aston University, Birmingham B4 7ET, UK}

\author{Nikita Tarasov}
\affiliation{Aston Institute of Photonics Technologies, Aston University, Birmingham B4 7ET, UK}

\author{Dmitry Churkin}
\affiliation{Aston Institute of Photonics Technologies, Aston University, Birmingham B4 7ET, UK}

\begin{abstract}
Loss of coherence with increasing excitation amplitudes and spatial size modulation is a fundamental problem in designing Raman fiber lasers. While it is known that ramping up laser pump power increases the amplitude of stochastic excitations, such higher energy inputs can also lead to a transition from a linearly stable coherent laminar regime to a non-desirable disordered turbulent state. This report presents a new statistical methodology, based on first passage statistics, that classifies lasing regimes in Raman fiber lasers, thereby leading to a fast and highly accurate identification of a strong instability leading to a laminar-turbulent phase transition through a self-consistently defined order parameter. The results have been consistent across a wide range of pump power values, heralding a breakthrough in the non-invasive analysis of fiber laser dynamics.
\end{abstract}

\pacs{42.65.Sf, 05.40.-a, 42.55.Ye, 42.55.Zz}
\maketitle

\section*{Introduction}

Raman fiber lasers are fascinating nonlinear optical systems which are conspicuous in their ability to give rise to a diverse range of operational regimes through an articulate control of system parameters \cite{turitsyn_nature_2013,turitsyn_adv_wave_turbulence,turitsyn_nature_2010,headley2004raman,agrawal2007,babin2007,churkin2011,turitsyna2012optical,churkin2010,Churkin2015}. These lasers utilize the stimulated Raman scattering mechanism occurring in the silica based fiber medium for gain. The small value of the Raman gain coefficient leads to the requirement of watt-order pump power levels and kilometer-order lengths of fiber for realizing the lasing threshold. The long cavity lengths give rise to very high mode densities, which together with the watt-order intracavity powers, lead to extremely complex nonlinear interactions between the modes. This has been explained within the framework of weak wave turbulence \cite{babin2007}. A complex interplay of nonlinearity and dispersion brings about a diverse palette of nonlinear dynamics, that include the generation of extreme events \cite{churkin2011}, formation of optical condensates \cite{turitsyna2012optical}, and a more recently discovered laser analog of a laminar to turbulent transition \cite{turitsyn_nature_2013}.

It has been reported \cite{turitsyn_adv_wave_turbulence} that unlike conventional modes of energy cascading, there is a certain phase locking between the inertial and dissipation energy scales at which energy cascades are respectively pumped in and out (dissipated) of the system, for which reason these two scales can not be non-equivocally unentangled. Effectively this amounts to an uncertainty in the inertial interval and hence in the categorization of the laminar-turbulent transition regime. While there are obvious theoretical and experimental challenges associated with such studies in relation to wave turbulent dynamics \cite{turitsyn_nature_2013}, an understanding of the underlying nonlinear dynamics of such many degree-of-freedom systems can lead to the realization of a new generation of high power, coherent laser systems. The parameter space is however extremely large, and identification of crossovers to different operational regimes - and more importantly, their predetermination is a problem of interest, especially from an application perspective.

In this study, we propose the use of an alternative time domain methodology for the classification of lasing regimes known in the statistical physics literature as persistence probability density function (PPDF) \cite{Bouchaud1990,current_science}, drawn from the rich legacy of first passage statistics which essentially quantifies how long a random walker \textit{persists} in any particular state before changing over to the other state(s). The effectiveness of the methodology is highlighted by using it to accurately identify the laminar-turbulent transition point in a Raman fiber laser. In the process, it is shown how the PPDF can be used for the definition of a suitable non-dimensional order parameter, which can then be used as a unique identifier of the lasing regime. 

\section*{Theory}

\subsection*{First passage statistics and the persistence probability distribution function}

The concept underlying {\it first passage extremal statistics}, also popularly referred to as {\it persistence statistics}, relies on a specific characterization of temporal statistics based on an analysis of two-point autocorrelation functions  \cite{Bouchaud1990}. The topic has been studied in grand details in the context of statistical mechanics \cite{current_science}, as also in a variety of interdisciplinary applications, like fractional dynamics \cite{Metzler2000,Metzler2014}, network characterization \cite{Lin2014}, immunological synapse modelling \cite{akc1,akc2,akc3} and gene data sequencing \cite{ChattopadhyayBI2015}. The key question here is the role that extremal statistic plays in the measurement of a statistical quantity. As an example, one may consider large amplitude fluctuations in a stochastic time series data set and estimate how such extremal events influence a mean statistical property e.g. the relevant temporal probability density function. Once characterized as a cumulative distribution function, the resulting power-law like statistics \cite{current_science} could be utilized to define an {\it order parameter} to represent phase crossovers or transitions \cite{bray1998}, and generically phase boundaries. 

This aforementioned order parameter that we intend to use for the purpose of characterizing our fiber laser data is defined through a two-step process, in which at step 1 we identify a local order parameter which then leads to a global order parameter at step 2. To arrive at these order parameters, we start from the following question: what is the probability $p(t_1; t_2)$ that the laser excitation intensity $X(t)$ changes sign $N$ times between $t_1$ and $t_2$ for $X(t) > X_0$? 

\begin{figure}[htbp]
        \includegraphics[width=0.49\textwidth]{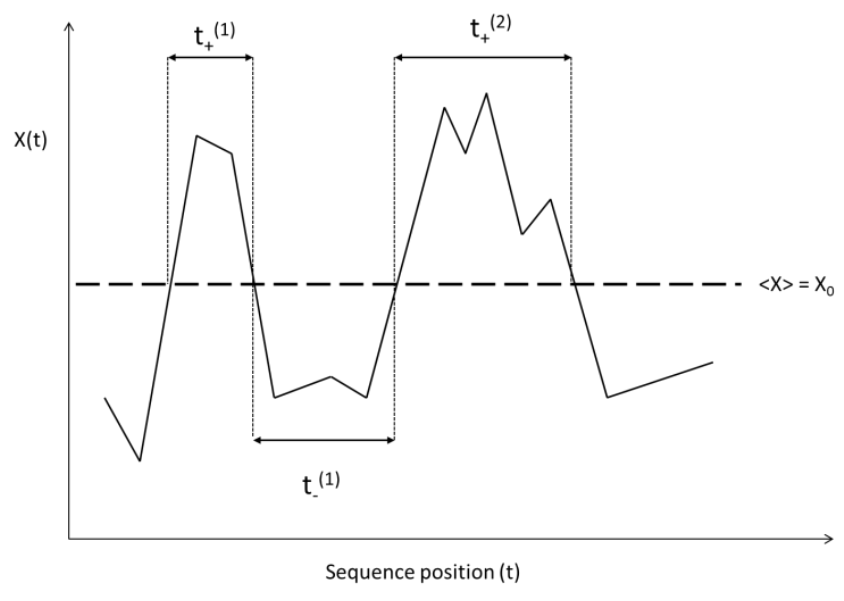}
        \caption{Brief Title Sentence: Cartoon diagram of Persistence mechanism
Legend: Stochastic dynamics of the fiber laser excitation amplitude $X(t)$ as a function of time $t$ across the threshold $X_0$, where $X_0$ represents the ensemble average of $X(t)$.}
        \label{Figure:fluctuations}
\end{figure}

Depending on the nature of random walk, the statistical physics parlance for the stochastic time series (fiber laser intensity) data, this probability is given by 
\begin{equation}
p(t_1,t_2)\sim \left(\frac{t_1}{t_2}\right)^{-\alpha_P}
\end{equation}
where the non-universal exponent $\alpha_P$ changes with the pump power $P$ and may actually be multi-fractal in nature. The subject has a rich heritage leading to the equation above that can be found in \cite{current_science, metzler2014first}. First passage statistics has been profitably utilized in interpreting phase transition scenarios \cite{metzler2014first}, a line of approach that we adopt here as well.

Depending on whether the statistics is evaluated above or below the line $X(t) = X_0$, the exponent $P$ would generally have two different values, one above the threshold and the other below it. In our following analysis, we would call the statistics above the threshold line $X(t) = X_0$ as being of the plus-type ($\alpha_P^+$) whereas the ones below the threshold will be of the minus-type ($\alpha_P^-$). The adjoining Figure 1 illustrates these definitions in which $t_+$ refers to the time spent between two crossings above the line $X(t) = X_0 (X(t) > X_0)$, whereas $t$ is the time spent between two crossings below the line $X(t) = X_0$. $\alpha_P$ is our initial estimate for a non-dimensional local order parameter at step 1. Our analysis uses $\alpha_P$ as the effective local order parameter characterizing the phase transition since it depicts the return time from a highly excited turbulent regime to that of a relatively transient laminar profile, thereby ensuring a much higher number of statistical events characterizing it due to low energy cost compared to the take-off time distribution scenario ( $\alpha_P^+$ statistics). $\alpha_P$ will shortly be shown to naturally lead to the ensemble averaged global order parameter.

\begin{figure}[htbp]
	\includegraphics[width=0.49\textwidth]{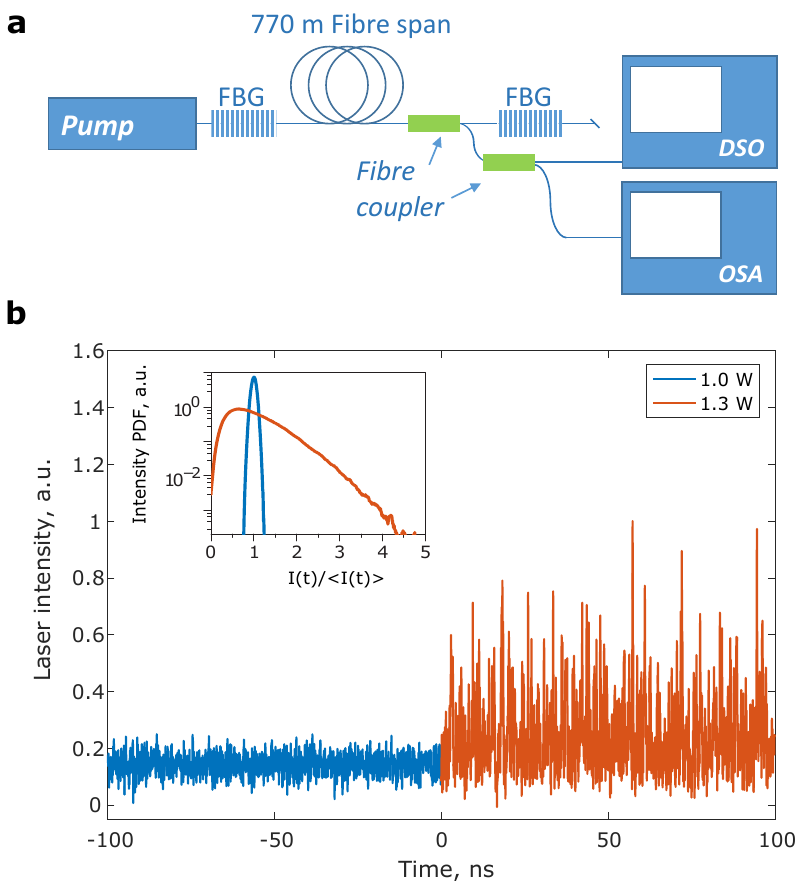}
	\caption{Brief Title Sentence: Experimental intensity time series data
Legend: \textbf{a.} Experimental configuration; FBG - fibre Bragg grating, DSO - digital storage oscilloscope, OSA - optical spectrum analyzer. \textbf{b.} Laminar (blue) and turbulent (orange) lasing regimes of the Raman fibre laser. Inset - Intensity PDFs of the corresponding regimes (also see Turitsyna et al \cite{turitsyn_nature_2013}).}
	\label{Figure:experimental_fluctuations}
\end{figure}

\section*{Results}

The PPDF methodology as described above is now applied to the study of the Raman fiber laser exhibiting a laminar-to-turbulent transition in intensity dynamics, as obtained in Turitsyna et al \cite{turitsyn_nature_2013}. The experimental apparatus uses a 770 meter long span of normal dispersion fiber (D = -44 ps/nm-km). Two highly reflective (R=0.99) 1 nm wide super-Gaussian gratings centered at 1555 nm were spliced to the ends of the fiber to form a Fabry-Perot cavity. This was then optically pumped using a watt-order Raman fiber laser operating in the 1455 nm region. A fiber-based fused coupler (99:1 ratio) was then used to obtain a small fraction of the intracavity radiation, which was interrogated using a 50 GHz optical photodetector, attached to a 36 GHz real-time digital storage oscilloscope. The intensity dynamics of the laser was then recorded over a time span equivalent to 120 round trips, sufficient to build up laser intensity statistics. The output characteristics were measured over a range of pump powers to progressively track the phase behavior with change in power inputs. Figure 2 shows the typical dynamics of a laminar-turbulent Raman fiber laser below (blue curves) and above (orange) the transition. In the laminar regime, the spectral width of emission, as monitored using an optical spectrum analyzer (resolution 20 pm), is $\sim$ 40 pm; this increases by almost an order of magnitude to $\sim$ 125 pm in the turbulent regime (also see Figure 4) \cite{turitsyn_nature_2013}. The spectral width remains stable over the course of measurement, with the changes in the instantaneous laser intensity primarily arising due to turbulent four wave mixing interactions between the different longitudinal modes \cite{babin2007}. The difference in the dynamics of the operational regimes are further evidenced by the well marked difference between their intensity probability distribution functions. Note that negative excursions in the measured intensities can arise from either photodetector noise, or electrical bandwidth limitations of the photodetector-oscilloscope. Nevertheless, the intensity statistics can be reliably estimated even under such noise and bandwidth limited considerations \cite{Gorbunov:14}. In our analysis, we tracked the range between $P$ = 0.85 W to 2.20 W in steps of 0.05 W. 

Figure 3 depicts three such persistence probability estimation within the aforementioned pump power value range in a log-log plot, where P refers to the value of the gradient at large $t$ values. It is important to note that for all pump powers, the corresponding PDFs show clear gradient change e.g. at around $log(t_+)$ in Figure 3. Our focus here is on the larger values of crossover times e.g. t to emphasize on the extremal values of excursion. The alluded $\alpha_P$ , though, estimates only the transient persistence of the data and not a global average over the given range of pump power values. In order to evaluate the average time persistence of the dynamics in either state, laminar or turbulent, we need an ensemble averaged global order parameter that will address the global symmetry of the model, a popular approach in liquid crystal studies \cite{degennes2002,prabirdaphysrep2015}. We define our effective (global) order parameter as
\begin{equation}
\beta_P = \int_{P_0}^{P}\frac{\partial\alpha_P^-}{\partial P}dP
\end{equation}
that quantifies the cumulative change of the gradient of the persistence distribution (measured by $\beta_P$) with change in pump power $P$ over the entire range of $\beta_P$ values. This is step 2 of the methodology in which a global order parameter $\beta_P$ is estimated that accurately quantifies the laminar-turbulent phase transition as shown in Figure 4. In our analysis, we have used the Ehrenfest classification of characterizing phase transitions based on discontinuous jumps of the order parameter \cite{amitzemansky}. In effect, $\beta_P$ is a non-Markovian order parameter that tracks all changes in the gradients of the persistence probability density function as a function of $P$. The emergence of $\beta_P$ is key to our analysis of the system as such a statistically averaged quantity reduces the impact of fluctuations, measured by the corresponding standard deviation, in the regression (least square) fits in Figure 3 that lead us to the values of $\alpha_P$, thereby stabilizing the phase transition picture overall. The error bars in Figure 4 bear testimony to this order parameter stability aspect.

\begin{figure}	
		\includegraphics[width=2.5in]{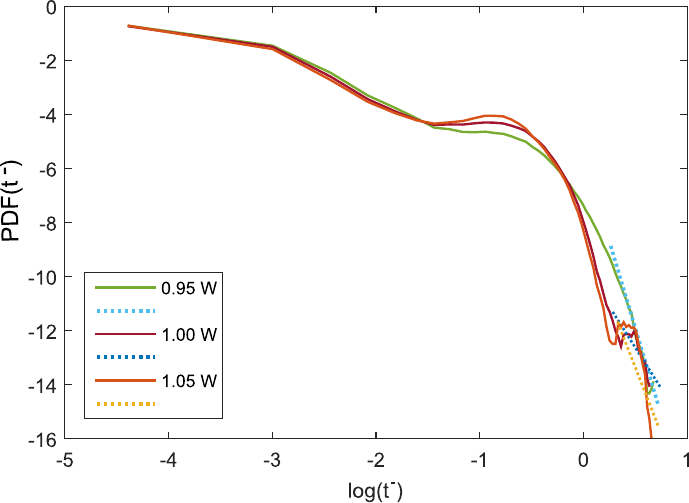}
		\caption{Brief Title Sentence: Temporal PDF plot
Legend: Probability Density Function (PDF) of \enquote{persistent} time series statistics in the log-log scale for pump power values $P$ = 0.95, 1.0 and 1.05 watts: {\bf minus}-type. The solid lines represent actual data while the dotted lines are the respective gradient fits (straight lines).}
		\label{PDF_fig}		
\end{figure}

Figure 4 outlines the strength of this algorithm by accurately matching the theoretical phase transition point (outset) with that observed in experiment as a step-like transition in the spectral width of the laser emission (inset). As the highlight of this study, this plot almost unerringly identifies a discontinuous jump at around $P$ = 1.05 W, the critical laminar-turbulent phase transition point that agrees well with the independent experimental result recently published in Figure 4 (inset) \cite{turitsyn_nature_2013}. The errorbars in the figure (outset), calculated as standard deviations of $\beta_P$ as evaluated from measured data, show fluctuations within 1\% of the estimated \enquote{order parameter} data, thereby confirming a very high degree of accuracy from this analysis. This is indeed remarkable given that the approach does not rely on any adjustable external parameter and relates only to the inherent stochasticity of the data leading to a self-consistent prediction of the laminar-turbulent phase transition point. 

\begin{figure}
	\includegraphics[width=2.8in,height=2.5in,angle=0]{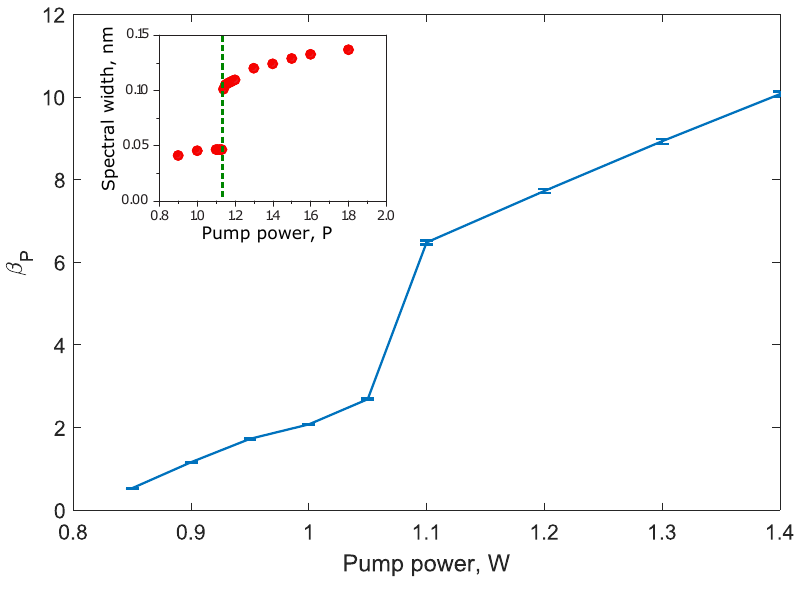}
	\caption{
	Brief Title Sentence: Phase transition from order parameter plot
Legend: Order parameter $\beta_P$ plotted against pump power $P$. Inset shows comparison with experimental data; theoretical phase transition prediction closely matches the experimental observation at $P \sim$ 1.05 watt.}
	\label{Figure:order_parameter}
\end{figure}

\section*{Conclusion}
This work presents the mapping of a well established analytical structure from statistical physics in analyzing a key photonics problem. The PPDF methodology exactly identifies the laminar-turbulent phase transition point from the  fiber laser intensity dynamics by estimating a self-consistent order parameter as a function of the (variable) excitation power used, independent of the nature of the sample. Indeed, the application of this technique is not limited to the case of the laminar-turbulent laser shown above, and can be used to study and classify the operational regimes of the wider class of quasi-continuous wave fiber lasers in general. The presented study extends further the scope of first passage statistics based analyzes in photonics as was reported in a recent work \cite{birkholz2015,erkintalo2015,Aragoneses,Kalashnikov2016}, where statistical methods were used to aid the understanding of the onset of rogue waves in certain optical systems. As a prime example of statistical physics based analysis of photonics data, this report holds the potential to serve as a much simpler, and yet a refined powerful diagnostic tool for the analysis of complex nonlinear fiber laser dynamics.

\section*{Acknowledgements}

AKC acknowledges \enquote{Aston University} funding for Gold Access of this article. SS, NT, and DVC acknowledge the EPSRC grant \enquote{ULTRALASER}. 

\section*{Author Contributions Statement}

AKC conceived the idea. ND wrote the numerical code for the processing of data. SS and NT did the experiments, supervised by DVC, AKC; AKC and SS jointly wrote the manuscript.

\section*{Ethical Approval and Informed Consent}
All experimental data (none involving live or biological samples) were collected in full compliance to requirements set by the Aston University Ethics Committee (https://www.ethics.aston.ac.uk/).

\end{document}